\documentclass[aps,prb,twocolumn,amsmath,amssymb]{revtex4-1}
\usepackage{graphicx}
\usepackage[dvipsnames,usenames]{xcolor}
\usepackage[colorlinks,citecolor=RoyalBlue]{hyperref}

\begin{document}
\title{Emergence of caustics in dynamics of the Kitaev model}
\author{Subhendu Saha$^1$}
\affiliation{$^1$Department of Physics, Presidency University \\ 86/1, College Street, Kolkata 700073, India}
\date{\today}
\begin{abstract}
We study quasiparticle dynamics in two-dimensional (2D) integrable Kitaev honeycomb model both without and in the presence of an external periodic drive. We identify light-cones in wavefunction propagation as a signature of quantum caustics, {\it i.e.} bright structures formed during quantum dynamics analogous to that of imperfect focusing in geometrical optics. We show that this dynamics follows an angle in spatial direction and it is anisotropic with respect to model parameters. Using coalescence of critical points, we provide an exact solution to the envelope of caustics, which corresponds to the Lieb-Robinson bound in 2D. Further, considering the system to be periodically driven, we point out that the caustics structure completely changes in presence of external time dependent drive.
\end{abstract}
\maketitle

\section{\label{Sec.1} Introduction}
Out of equilibrium dynamics of closed quantum many-body systems have been a field of study for posing different long-standing problems in physics \cite{eisert15}. An important feature of this dynamics is the existance of light-cone like structures within \emph{Lieb-Robinson bound} \cite{lieb72} which is the quantum speed limit of information propagation for such systems. All the informations beyond that limit is exponentially suppressed and correlation functions significantly grow in time only within that light-cone \cite{hastings06,bravyi06}. In a recent work, it is shown that this light-cone holds key aspects of local dynamics followed by a quench \cite{calabrese06} which creates excitations propagation through the system with a finite group velocity. This offers a promising potential to investigate non-equilibrium dynamics \cite{calabrese05,kollath07,laeuchli07, barmet12,cramer08,hauke13,eisert13,manmana09} with analytical and numerical formalisms. In addition to being theoretically appealing, these questions are strongly motivated by the rapid experimental progress in realization of optical lattice systems using Bose-Hubbard \cite{greiner02,trotzky12}, spin \cite{scha13,jurcevic14,langen13,fuku13,simon11,kim10, bonnes14}, and quasi-one dimensional \cite{kino06,hoffer07} models with ultra-cold atoms, molecules and trapped ions. The ability to address individual sites \cite{weit11,cheneau12,sherson10,bakr09} by accurate control enables unique state preparation and measurement of local observables. Moreover, long coherence times \cite{blatt12} of continuous atomic systems \cite{geiger14} make them useful while studying dynamics. Light-cone effects now have been observed in all sorts of systems mentioned above \cite{kajala11,jurcevic14,cheneau12,richerme14,bonnes14}.

\par Recent years have witnessed substantial progress in understanding the dynamics of periodically driven (Floquet) systems. Such well controlled periodic drive opens up new possibilities for quantum simulation using ultra-cold atoms. Consider a system with Hamiltonian being a periodic function of time $H(t)=H(t+T)$ with a period $T>0$. The eignstates of such systems constantly change with time, but the long time dynamics can be studied from corresponding \emph{Floquet operator} $U(T)$ \cite{arimondo12}, which is the time evolution operator for one time period of driving. Therefore, one can define an effective Hamiltonian from $U(T)=\exp[-iH_{eff}T]$. The time evolution of the system after any integer number of period may be completely described by $H_{eff}$, which can be tuned using suitable driving protocol via model parameters. This allows to realize topological models using shaken optical lattice \cite{jotzu14,hauke12}, engineering of artificial gauge fields \cite{goldman14,jaksch03,struck12,struck13,aidel15}, new topological materials \cite{goldman16,moessner17} and several other interesting phenomenon involving periodic drive \cite{rudner13,dalessio14,bukov15,eckardt17,weinberg17}.

\par In geometrical optics \emph{caustics} are the locus of infinite ray density where singularity arises due to the divergence of light intensity. They are the example of catastrophes where classical ray theory fails and one must appeal to wave theory of light in order to have the full description. The phase singulariy in classical waves is however removed by quantization of field excitations leading to \emph{quantum catastrophe} \cite{ulf02,berry04,berry08}. Caustics may also appear during the dynamics of quantum many-particle system when singularity in Fock space is removed by the discreteness in excitation number \cite{odell12}. Several experiments with cold atoms have observed caustics in trajectories of atoms trapped in a linear magnectic field \cite{rooijakkers03}, dynamics of BEC in an optical lattice \cite{huckans09}, atomic cloud reflected from an optical barrier in presence of gravity \cite{rosenblum14}. On theoretical side, caustics are predicted for expanding condensate in an optical lattice \cite{chalker09}, dynamics of bosonic Josephson junctions in two lattice sites \cite{odell12} involving experiments of a.c. and d.c. Josephson effects \cite{levy07}, self-trapping \cite{albiez05}, matter-wave interference in double well \cite{schumm05} with BEC. Although it is not mentioned by the authors, cautics can be seen in figures of the papers on collapse and spreading of a polariton condensate using ultra-fast laser pulse \cite{dominici15}, quantum walk of interacting bosons in an optical lattice \cite{preiss15}, density modulation in supersonic BEC against an obstacle \cite{carusotto06}. In a recent study, it is shown that catastrophe exists in many-body wavefunctions following a quench in two-mode Bose-Hubbard model \cite{mumford17} and light-cone like structures formed during quasiparticle propagation can be thought of as quantum caustics for Ising model in transverse field \cite{kirkby17}.

\par Motivated by this kind of works we make a move from previous one-dimensional (1D) systems to a two-dimensional (2D) integrable system. For that, we shall mostly follow the techniques in Ref.~\cite{kirkbythesis17}. Our primary interest is to see similar kind of light-cone like structure in 2D. We take spin-1/2 model of 2D honeycomb lattice described by Kitaev Hamiltonian. The main results that we obtain from our study are following. First, we develop an analytical framework of studying single quasiparticle dynamics for Kitaev model in terms of catastrophe function. Second, we numerically show that excitations travel in a particular direction in lattice plane and gradually decreases with respect to a model parameter. Thus that follows a spatial anisotropy which we address by the amplitude variation of wavefunction on some defined angle between the parameters. Third, we calculate the exact analytic expressions for caustics trajectories which gives the Leib-Robinson bound in 2D. Fourth, we periodcally vary a parameter of the Hamiltonian with a time evolution operator $U$ and show that the caustics no longer follow a light-cone. 

\par The rest of the paper is planned as follows. In Sec.~\ref{Sec.2} we discuss some of the properties of the Kitaev honeycomb model, its energy dispersion and phases. In Sec.~\ref{Sec.2.a} we study the angle dependence on quasiparticle dynamics. This is followed by Sec.~\ref{Sec.2.b} where we demonstrate the positions of quantum caustics in a light-cone like structure bounded by Lieb-Robinson limit. Next, in Sec.~\ref{Sec.3} we introduce periodic drive protocol and see its effects on caustics. Finally, we summarize our results in Sec.~\ref{Sec.4} with some concluding remarks.

\section{\label{Sec.2} Kitaev honeycomb model}
The Kitaev model is a fermionic spin model with spin-1/2's being placed on the sites of a honeycomb lattice following a Hamiltonian \cite{kitaev06} of the form
\begin{equation}\label{Eq.2.1}
H=\sum_{j+l=even}\left(J_{1} \sigma_{j,l}^{x}\sigma_{j+1,l}^{x}+J_{2} \sigma_{j-1,l}^{y}\sigma_{j,l}^{y}+J_{3} \sigma_{j,l}^{z}\sigma_{j,l+1}^{z}\right)
\end{equation}
where $j$ and $l$ represent the column and row indices of the lattice, $\sigma_{m,n}^{\alpha},\,\alpha\in\{x,y,z\}$ are Pauli matrics at site labeled by (m,n), and $J_{1}$, $J_{2}$ and $J_{3}$ are coupling parameters. From now on, we assume that all couplings are time indipendent and $J_{i}\geq 0$.
\begin{figure}[h]
\includegraphics[width=\columnwidth]{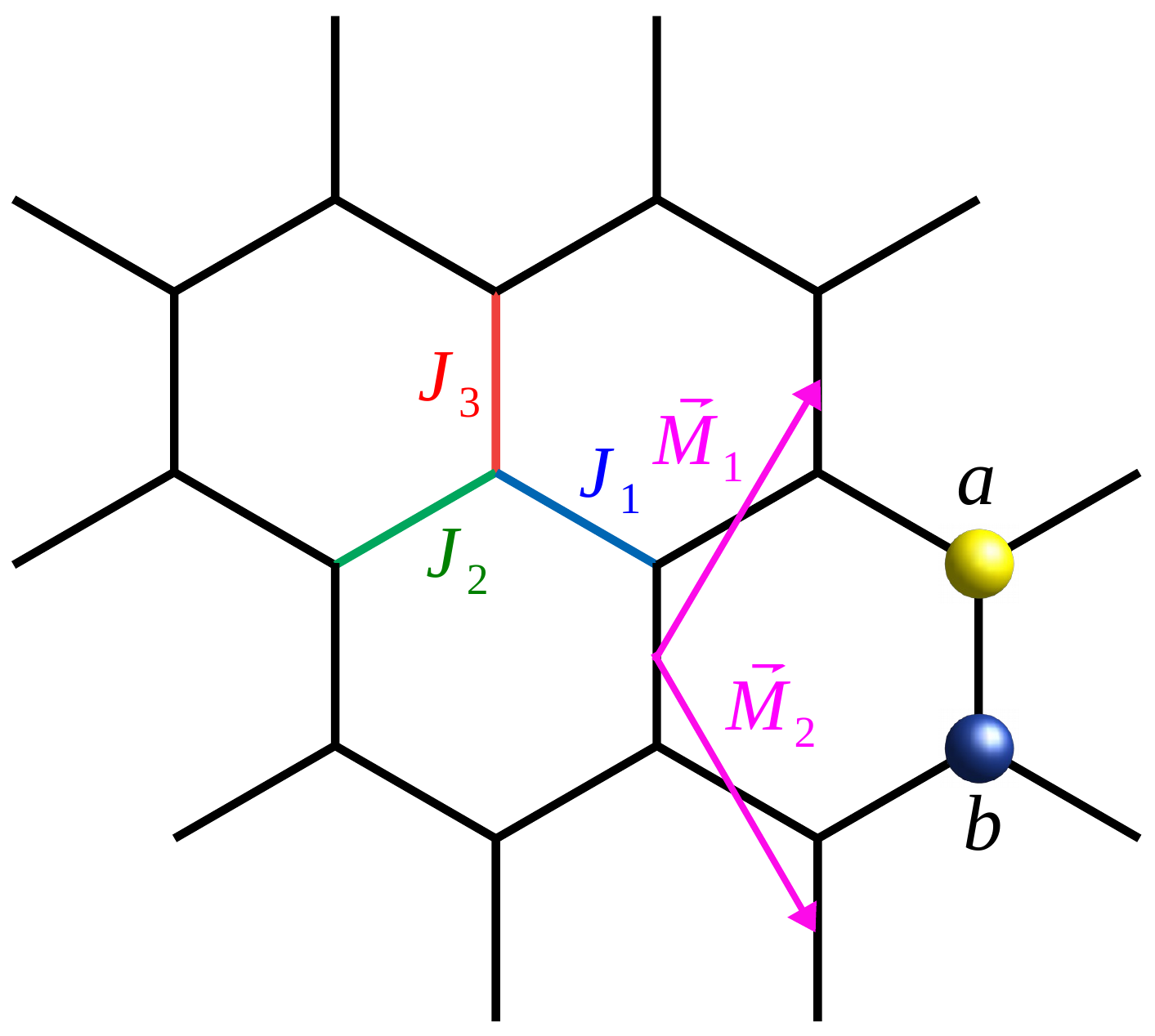}
\caption{(Color online) Kitaev honeycomb lattice with couplings $J_{1}$, $J_{2}$ and $J_{3}$. $\mathbf{M}_{1}$ and $\mathbf{M}_{2}$ are the spanning vectors of the lattice, and $a$ and $b$ represent two inequivalent sites of a unit cell.}
\label{Fig.2.1}
\end{figure} 
A schematic diagram of the honeycomb lattice is in Fig.~\ref{Fig.2.1} showing the bonds with $J_{1}$, $J_{2}$ and $J_{3}$. To denote the unit cell of the lattice, we label the two sites of a vertical bond with $a$ and $b$. These two positions have $j+l$ equal to odd or even integers respectively. If there are $N$ (assumed to be even) number of lattice sites, the number of unit cells is $N/2$. It is convenient to set the nearest neighbour distance to be $1/\sqrt{3}$, so that, each unit cell is labeled by a vector $\mathbf{n}=\hat{i}n_{1}+\left(\frac{1}{2}\hat{j}+\frac{\sqrt{3}}{2}\hat{i}\right)n_{2}$, where $n_{1}$, $n_{2}$ run over all integers and $\mathbf{M}_{1}=\frac{1}{2}\hat{i}+\frac{\sqrt{3}}{2}\hat{j}$ and $\mathbf{M}_{2}=\frac{1}{2}\hat{i}-\frac{\sqrt{3}}{2}\hat{j}$ are the spanning vectors which join neighbouring unit cells in the  reciprocal lattice.
\par One of the main properties that makes Kitaev model thoretically fascinating is that, one can map it onto a non-interacting fermionic model using suitable Jordan-Wigner transformation \cite{feng07,lee07,chen08,nussinov08}. The Hamiltonian takes the form
\begin{equation} \label{Eq.2.2}
H_F=i\sum_{\mathbf{n}}\left(J_{1}b_{\mathbf{n}}a_{\mathbf{n}-\mathbf{M}_{1}}+J_{2}b_{\mathbf{n}}a_{\mathbf{n}+\mathbf{M}_{2}}+J_{3}b_{\mathbf{n}}a_{\mathbf{n}}D_{\mathbf{n}}\right)
\end{equation}
where $a_{\mathbf{n}}$ and $b_{\mathbf{n}}$ are Majorana fermions sitting at the top and bottom sites, defined by Majorana operators
\begin{equation} \label{Eq.2.3}
\begin{aligned}
\hat{a}_{j,l}&=\left(\prod_{i=-\infty}^{j-1}\sigma_{i,l}^{z}\right)\sigma_{j,l}^{y}  \quad \text{for} \quad j+l=even \\
\hat{b}_{j,l}&=\left(\prod_{i=-\infty}^{j-1}\sigma_{i,l}^{z}\right)\sigma_{j,l}^{x} \quad \text{for} \quad j+l=odd
\end{aligned}
\end{equation}
These operators are Hermitian and satisfy anticommutation relations $\{\hat{a}_{m,n},\hat{a}_{m^{\prime},n^{\prime}}\}=\{\hat{b}_{m,n},\hat{b}_{m^{\prime},n^{\prime}}\}=2\delta_{m,m^{\prime}}\delta_{n,n^{\prime}}$ and $\{\hat{a}_{m,n},\hat{b}_{m^{\prime},n^{\prime}}\}=0$. Also the operator $\hat{D}_{\mathbf{n}}$ takes the eignvalues $\pm 1$ independently for each $\mathbf{n}$, therefore $2^{N}$ dimensional Hilbert space decomposes into $2^{N/2}$ sectors. The ground state of the model corresponds to $\hat{D}_{\mathbf{n}}=1$ on all the bonds irrespective of the sign of $J_{3}$ due to special symmetry of the model. Also $\hat{D}_{\mathbf{n}}$ being the constant of motion, dynamics of the model never takes the system outside $\hat{D}_{\mathbf{n}}=1$. Throughout our calculation we will be within that sector only.
\par The Fourier transform of the Majorana operators are defined as 
\begin{equation} \label{Eq.2.4}
\begin{aligned}
\hat{a}_{\mathbf{n}}&=\sqrt{\frac{4}{N}}\sum_{\mathbf{k}\in\frac{1}{2}BZ}\left(\hat{a}_{\mathbf{k}}e^{i\mathbf{k}.\mathbf{n}}+\hat{a}^{\dagger}_{\mathbf{k}}e^{-i\mathbf{k}.\mathbf{n}}\right) \\
\hat{b}_{\mathbf{n}}&=\sqrt{\frac{4}{N}}\sum_{\mathbf{k}\in\frac{1}{2}BZ}\left(\hat{b}_{\mathbf{k}}e^{i\mathbf{k}.\mathbf{n}}+\hat{b}^{\dagger}_{\mathbf{k}}e^{-i\mathbf{k}.\mathbf{n}}\right)
\end{aligned}
\end{equation}
where the operators also follow the anticommutation relations $\{\hat{a}_{\mathbf{k}},\hat{a}^{\dagger}_{\mathbf{k}^{\prime}}\}=\{\hat{b}_{\mathbf{k}},\hat{b}^{\dagger}_{\mathbf{k}^{\prime}}\}=\delta_{\mathbf{k}\mathbf{k}^{\prime}}$. Here the summation goes over half Brillouin zone (BZ), for this model a convenient choice is rhombus whose vertices lie at $(k_{x},k_{y})=(\pm 2\pi/\sqrt{3},0)$ and $(0,\pm 2\pi/3)$. \\
For $\hat{D}_{\mathbf{n}}=1$, the Hamiltonian in Eq.~(\ref{Eq.2.2}) can be written in terms of Fourier modes 
\begin{equation} \label{Eq.2.5}
\begin{aligned}
H&=\sum_{\mathbf{k}\in\frac{1}{2}BZ}\psi^{\dagger}_{\mathbf{k}}H_{\mathbf{k}}\psi_{\mathbf{k}} \quad \text{where} \quad \psi_{\mathbf{k}}=\left({\begin{array}{c} a_{\mathbf{k}} \\ b_{\mathbf{k}} \end{array}} \right) \\
H_{\mathbf{k}}&=2\left[J_{1}\sin(\mathbf{k}.\mathbf{M}_{1})-J_{2}\sin(\mathbf{k}.\mathbf{M}_{2})\right]\tau_{1} \\
&+ 2\left[J_{3}+J_{1}\cos(\mathbf{k}.\mathbf{M}_{1})+J_{2}\cos(\mathbf{k}.\mathbf{M}_{2})\right]\tau_{2}
\end{aligned}
\end{equation}
where $\tau_{\alpha},\alpha\in\{1,2,3\}$ are the Pauli matrices denoting psudospin. We carry out a global rotation $U=\exp[-i\frac{\pi}{4}\tau_{1}]$ so that $\psi \rightarrow U^{\dagger} \psi^{\prime}$ and $H_{\mathbf{k}}^{\prime}=UH_{\mathbf{k}}U^{\dagger}$. Now we write the Hamiltonian in the following form \cite{sengupta08,mondal08}
\begin{equation}
\begin{aligned}
H&=\sum_{\mathbf{k}\in\frac{1}{2}BZ}{\psi^{\prime}_{\mathbf{k}}}^{\dagger}H^{\prime}_{\mathbf{k}}\psi^{\prime}_{\mathbf{k}} \\
H^{\prime}_{\mathbf{k}}&=\epsilon_{\mathbf{k}}\tau_{3}+\Delta_{\mathbf{k}}\tau_{1}
\end{aligned}
\end{equation}
The energy spectrum hence consists of two bands with energies
\begin{equation} \label{Eq.2.6}
\begin{aligned}
E_{\mathbf{k}}^{\pm}=&\pm2[\{J_{1}\sin(\mathbf{k}.\mathbf{M}_{1})-J_{2}\sin(\mathbf{k}.\mathbf{M}_{2})\}^{2} \\
&+
\{J_{3}+J_{1}\cos(\mathbf{k}.\mathbf{M}_{1})+J_{2}\cos(\mathbf{k}.\mathbf{M}_{2})\}^{2}]^{1/2}
\end{aligned}
\end{equation}
For $|J_{1}-J_{2}|\leq J_{3}\leq J_{1}+J_{2}$ the energy bands touch each other, $E_{\mathbf{k}}=0$  and the gap $\Delta_{\mathbf{k}}=E^{+}_{\mathbf{k}}-E^{-}_{\mathbf{k}}$ vanishes for special values of $\mathbf{k}$ in that phase.

\subsection{\label{Sec.2.a} Dynamics of a single quasiparticle}
It turns out that it is possible to write down the Kitaev Hamiltonian in a similar language as that is used for 1D Ising model. Although this choice is not strictly necessary for the analysis but we will use it to make us familiar with Ref.~\cite{kirkbythesis17}. Let us define a set of complex fermions on each link of the vertical bonds \cite{chen08}
\begin{equation} \label{Eq.2.7}
\begin{aligned}
d_{\mathbf{n}}&=\left(b_{\mathbf{n}}+ia_{\mathbf{n}}\right)/2 \\
d^{\dagger}_{\mathbf{n}}&=\left(b_{\mathbf{n}}-ia_{\mathbf{n}}\right)/2
\end{aligned}
\end{equation}
We redefine the Hamiltonian in terms of these fermions as $H=\sum_{\mathbf{k}\in \frac{1}{2}BZ}\bar{\psi}^{\dagger}_{\mathbf{k}}H^{\prime}_{\mathbf{k}}\bar{\psi}_{\mathbf{k}}$ with ${\bar{\psi}_{\mathbf{k}}}=({\begin{array}{c} d_{\mathbf{k}}, d^{\dagger}_{-\mathbf{k}}\end{array}})^{T}$. Then diagonalizing this Hamiltonian in terms of Bogoliubov quasiparticle defined as $\gamma_{\mathbf{k}}=u_{\mathbf{k}}d_{\mathbf{k}}+v_{\mathbf{k}}d^{\dagger}_{-\mathbf{k}}$ leads to $H=\sum_{\mathbf{k}\in \frac{1}{2}BZ} E_{\mathbf{k}}\left(\tilde{\gamma}^{\dagger}_{\mathbf{k}}\tilde{\gamma}_{\mathbf{k}}-\frac{1}{2}\right)$. Here $\tilde{\gamma}^{(\dagger)}_{\mathbf{k}}$ is the annhilation (creation) operators for Bogoliubov modes with momentum $\mathbf{k}$. The dynamics of a quasiparticle at a single site is then given by
\begin{equation} \label{Eq.2.8}
|\psi(t)\rangle=e^{-i\hat{H}t/\hbar}\,\hat{\gamma}^{\dagger}_{r=0}|0\rangle=\frac{e^{i\theta(t)}}{\sqrt{N}}\sum_{\mathbf{k}}e^{-iE_{\mathbf{k}}t/\hbar}|\mathbf{k}\rangle
\end{equation}
where $\hat{\gamma}^{(\dagger)}_{r}$ annhilates (creates) a quasiparticle at site $r$, $\theta(t)$ denotes the unobservable global phase factor of the Hamiltonian and $E_{\mathbf{k}}$ is the dispersion relation of the model. One can consider a state in Eq.~(\ref{Eq.2.8}) projected onto real space, $\psi(\mathbf{r},t)\equiv\langle\mathbf{r}|\psi(t)\rangle$ to have the wavefunction
\begin{equation} \label{Eq.2.9}
\psi(\mathbf{r},t)=\frac{e^{i\theta(t)}}{N}\sum_{\mathbf{k}\in \frac{1}{2}BZ}e^{i\mathcal{S}(\mathbf{k};\mathbf{r},t)}
\end{equation}
where the functional $\mathcal{S}(\mathbf{k};\mathbf{r},t)=-(\mathbf{k}.\mathbf{r}+2E_{\mathbf{k}}t)$ and $\theta(t)=t/2\sum_{\mathbf{k}}E_{\mathbf{k}}$. To determine the caustics we further define 
\begin{equation} 
\begin{aligned}
k_{1}&=\mathbf{k}.\mathbf{M}_{1}=\frac{\sqrt{3}}{2}k_{x}+\frac{3}{2}k_{y} \\
k_{2}&=\mathbf{k}.\mathbf{M}_{2}=\frac{\sqrt{3}}{2}k_{x}-\frac{3}{2}k_{y}
\end{aligned}
\end{equation}
where $k_{1}$, $k_{2}$ are in units of the lattice constant $a$. Thus one can write
\begin{equation} \label{Eq.2.10}
\begin{aligned}
\mathbf{k}.\mathbf{r}&=\frac{1}{\sqrt{3}}(k_{1}+k_{2})x+\frac{1}{3}(k_{1}-k_{2})y \\
&=k_{1}x_{1}+k_{2}x_{2},
\end{aligned}
\end{equation}
where $x_{1}=\frac{x}{\sqrt{3}}+\frac{y}{3}$ and $x_{2}=\frac{x}{\sqrt{3}}-\frac{y}{3}$ respectively. In Sec.~\ref{Sec.2} we have labeled the unit cell with $\mathbf{n}$, so it is instructive to express all spatial coordinates (in our case $x_{1}$, $x_{2}$) in terms of $n_{1}$ and $n_{2}$. 
\begin{figure}[h]
\includegraphics[width=\columnwidth]{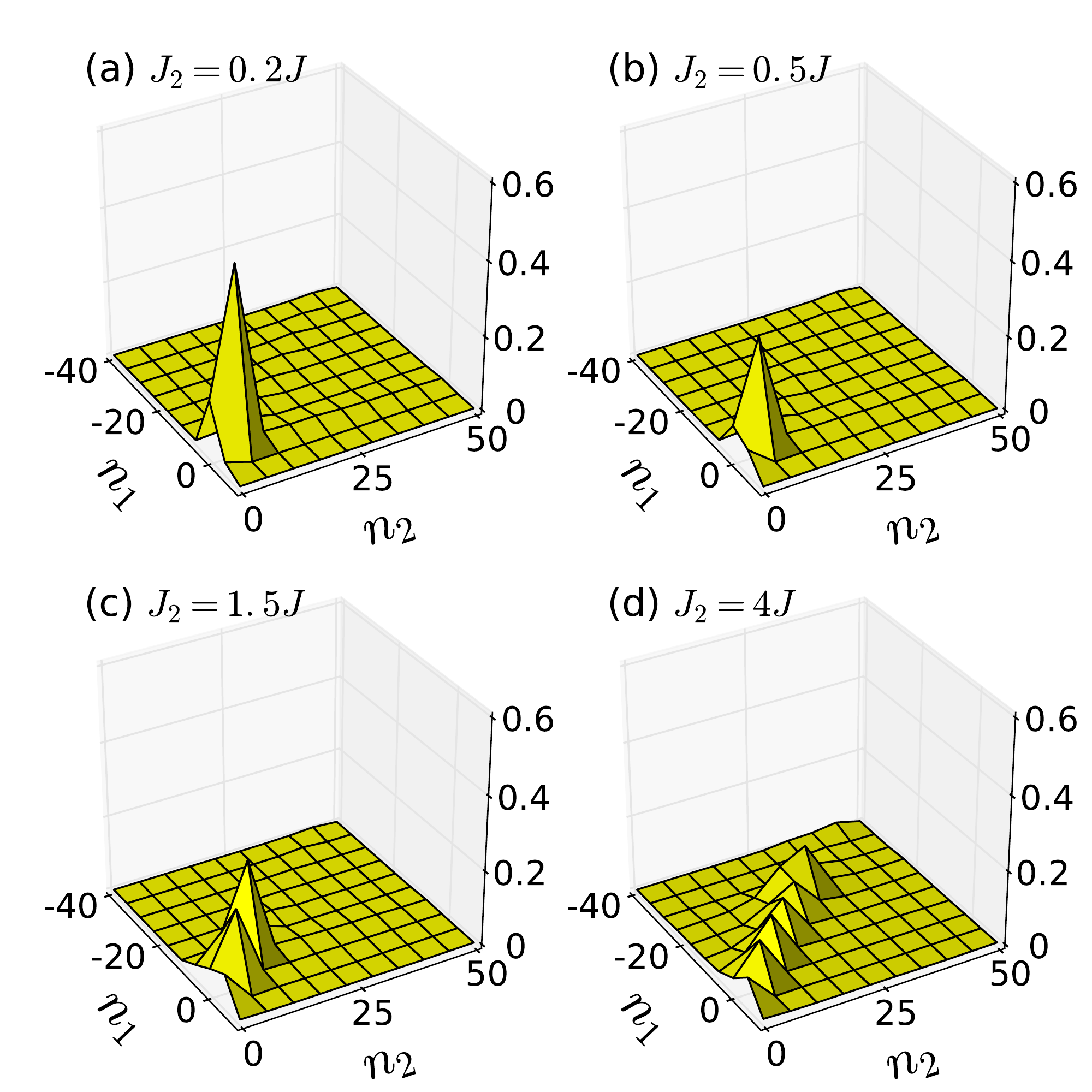}
\caption{(Color online) Plots of $|\psi(\mathbf{r},t)|$ as a function of $\mathbf{r}$ for several representative values of $J_{2}/J$ for $J_{1}=J$ and $J_{3}=5$ at $t=3$. The plot displays the change in wavefunction amplitude as a function of $J_{2}/J_{1}$.}
\label{Fig.2.2}
\end{figure} 
We numerically calculate the amplitude of the discrete wavefunction in Eq.~(\ref{Eq.2.9}) as a function of $n_{1}$ and $n_{2}$ for a fixed time, where $x=\sqrt{3}(n_{1}+n_{2}/2)$ and $y=3n_{2}/2$ . A plot in Fig.~\ref{Fig.2.2} shows $|\psi(\mathbf{r},t)|$ for several representive values of $J_{2}/J$ for a fixed $J_{1}=J$ and $J_{3}=5$ at $t=3$. We find that $|\psi(\mathbf{r},t)|$ takes a direction in the $n_{1}-n_{2}$ plane as $J_{2}/J_{1}$ ratio is increased. This clearly shows that the wavefunction is highly anisotropic in this limit. The change in amplitude can also be seen as we go from the limit $J_{2}\ll J_{1}$ to the limit $J_{2}\gg J_{1}$. 
\begin{figure}[h]
\includegraphics[width=\columnwidth]{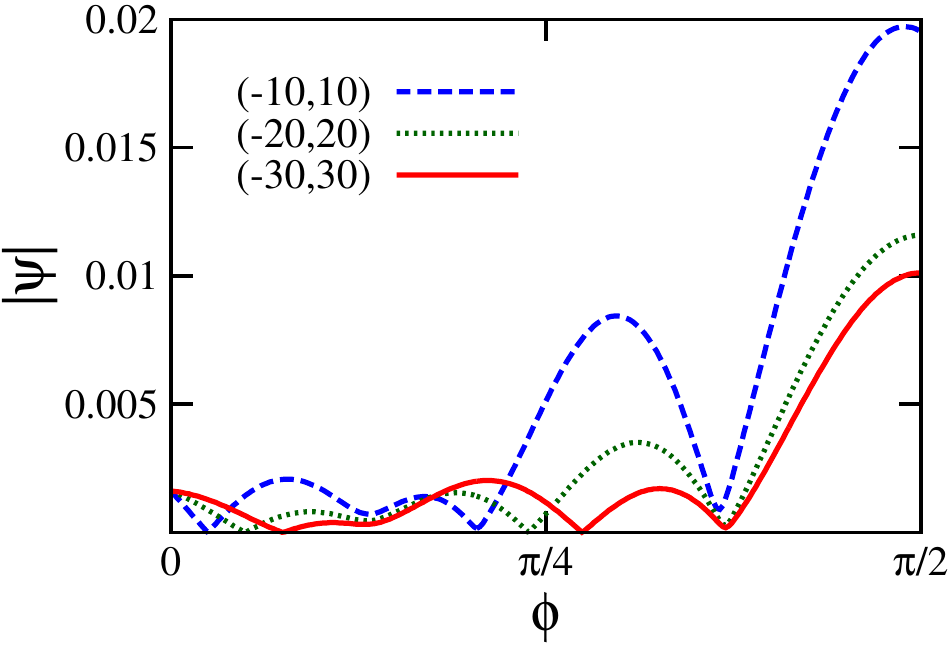}
\caption{(Color online) Plot of $|\psi(\mathbf{r},t)|$ at points $(-10, 10), (-20, 20), (-30, 30)$ along $-45^{\circ}$ in the $n_{1}-n_{2}$ plane as a function of $\phi=\tan^{-1}(J_{2}/J_{1})$ keeping $J^{2}=1$ fixed.}
\label{Fig.2.3}
\end{figure}
\par We do a more detail analysis of this spatial anisotropy of wavefunction as a function of $J_{2}/J_{1}$. We define a parameter $\phi$ such that $J_{1[2]}=J\,\cos\,\phi[\sin\,\phi]$. A variation in $\phi$ thus changes the ratio $J_{2}/J_{1}$ from $0$ to $\infty$ while keeping $J_{1}^{2}+J_{2}^{2}=J^{2}=1$ fixed. The plot of $|\psi(\mathbf{r},t)|$ at three points $(n_{1}, n_{2})=(-10, 10), (-20, 20), (-30, 30)$ along the $-45^{\circ}$ line in the $n_{1}-n_{2}$ plane as a function of $\phi$ shown in Fig.~\ref{Fig.2.3} depicts the nature of spatial anisotropy. We see that as the ratio of $J_{2}/J_{1}$ is varied from $0$ to $\infty$, the amplitude of wavefunction becomes maximum when $J_{2}\gg J_{1}$ ($\phi=\pi/2$). Further the amplitude decreases as one goes far along that line, as expected from Fig.~\ref{Fig.2.2}. This allows us to conclude that spatial anisotropy of the wavefunction amplitude depends significantly on the ratio of $J_{2}/J_{1}$.

\subsection{\label{Sec.2.b} Position of the caustics}
Locating the caustics from its generating function (action) would require minimizing the effective action $\mathcal{S}$ with respect to all momentum values and finding the saddles at a critical line. This can be done by satisfying the following conditions simultaneously
\begin{equation} \label{Eq.2.11}
\frac{\partial \mathcal{S}}{\partial k_{1}}=\frac{\partial \mathcal{S}}{\partial k_{2}}=0 \quad \text{and} \quad \frac{\partial^{2} \mathcal{S}}{\partial k_{1}^{2}}\, \frac{\partial^{2} \mathcal{S}}{\partial k_{2}^{2}}-\left(\frac{\partial^{2} \mathcal{S}}{\partial k_{1} \partial{k_{2}}}\right)^{2}=0
\end{equation}
We define two parameters $\alpha=J_{1}/J_{3}$ and $\beta=J_{2}/J_{3}$ such that Eq.~(\ref{Eq.2.11}) together with $\mathcal{S}(\mathbf{k};\mathbf{r},t)=-(k_{1}x_{1}+k_{2}x_{2}+2E_{\mathbf{k}}t)$ yield three equations
\begin{equation} \label{Eq.2.12}
x_{1}=\frac{J_{3}t}{E^{\prime}_{\mathbf{k}}}\,C_{1\mathbf{k}}, \quad x_{2}=\frac{J_{3}t}{E^{\prime}_{\mathbf{k}}}\,C_{2\mathbf{k}}
\end{equation}
\begin{equation} \label{Eq.2.13}
\begin{aligned}
\left(4{E^{\prime}_{\mathbf{k}}}^{2} D_{1\mathbf{k}}+C_{1\mathbf{k}}^{2}\right)&\left(4{E^{\prime}_{\mathbf{k}}}^{2} D_{2\mathbf{k}}+C_{2\mathbf{k}}^{2}\right) \\
&=\left[C_{1\mathbf{k}}C_{2\mathbf{k}}+4\alpha\beta{E^{\prime}_{\mathbf{k}}}^{2}\cos (k_{1}+k_{2})\right]^{2}
\end{aligned}
\end{equation}	
where
\begin{equation}
\begin{aligned}
C_{1\mathbf{k}}&=2\alpha\sin k_{1}+2\alpha\beta\sin (k_{1}+k_{2}) \\
C_{2\mathbf{k}}&=2\beta\sin k_{2}+2\alpha\beta\sin (k_{1}+k_{2}) \\
D_{1\mathbf{k}}&=\alpha\cos k_{1}+\alpha\beta\cos (k_{1}+k_{2}) \\
D_{2\mathbf{k}}&=\beta\cos k_{2}+\alpha\beta\cos (k_{1}+k_{2})
\end{aligned} 
\end{equation}
and
\begin{equation}
E_{\mathbf{k}}^{\prime}=\left[\left(1+\alpha\cos k_{1}+\beta\cos k_{2}\right)^{2}+\left(\alpha\sin k_{1}-\beta\sin k_{2}\right)^{2}\right]^{1/2}
\end{equation}
\begin{figure}[h]
\includegraphics[width=\columnwidth]{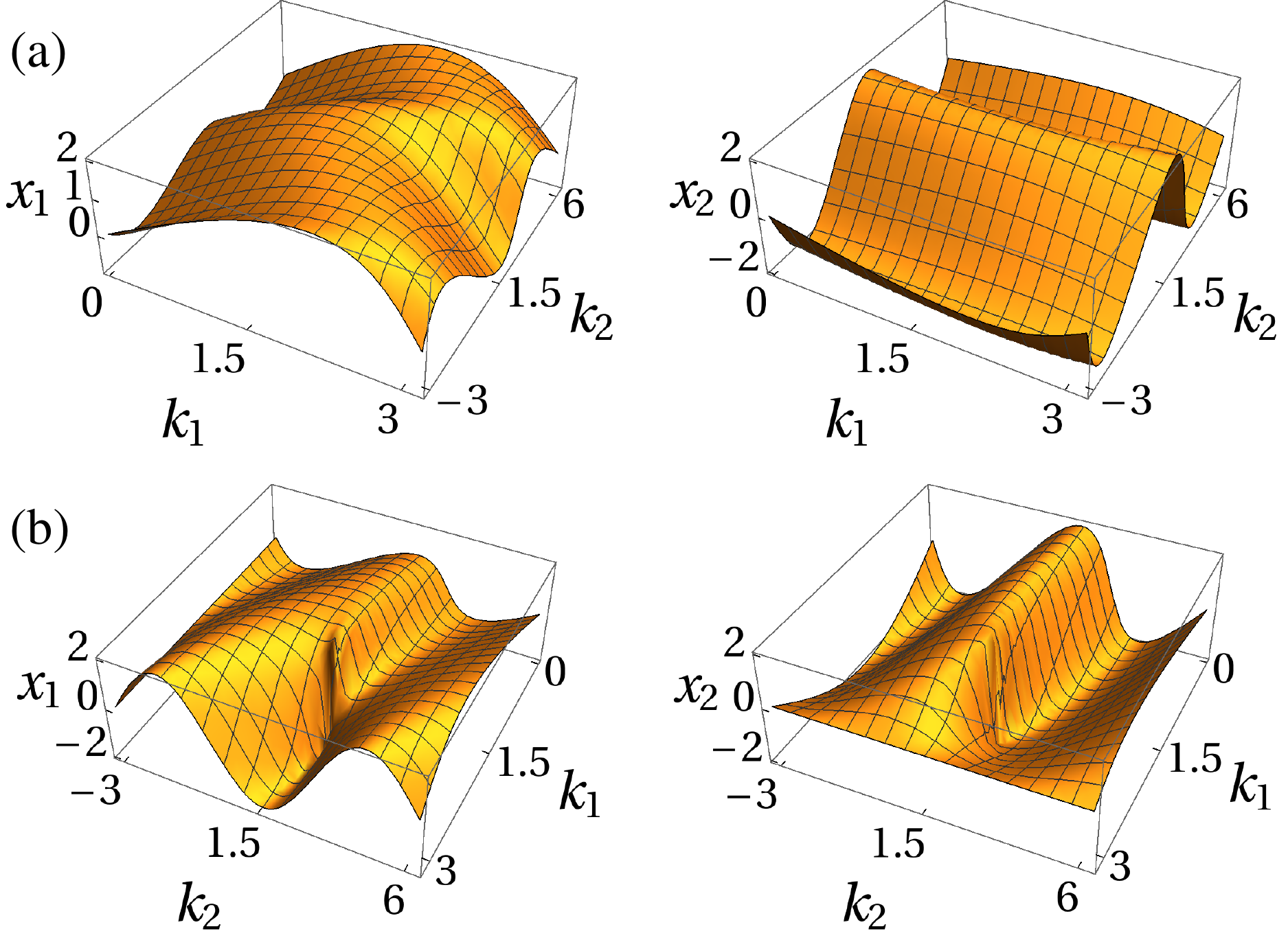}
\caption{(Color online) Relation between $x_{1}$, $x_{2}$ and $t$ representing the caustics trajectories in (a) gapped (b) gapless phase}
\label{Fig.2.4}
\end{figure}
We numerically solve Eq.~(\ref{Eq.2.12}) and Eq.~(\ref{Eq.2.13}) and plot the trajectories corresponding to the caustics in Fig.~\ref{Fig.2.4} for gapped and gapless phases of the model. We further attempt to solve this analytically for those two limiting cases. At $J_{3}\gg J_{1}, J_{2}$ (ie. $\alpha, \beta\ll 1$) in the gapped phase Eq.~(\ref{Eq.2.12}) has leading orders $J_{1}\sin k_{1}=x_{1}/2t$ and $J_{2}\sin k_{2}=x_{2}/2t$ which have the solutions $k_{1}=\sin^{-1}(x_{1}/2J_{1}t)$ and $k_{2}=\sin^{-1}(x_{2}/2J_{2}t)$. One can clearly see that $k_{1}$ and $k_{2}$ have no solution for $x_{1}>2J_{1}t$ and $x_{2}>2J_{2}t$. Thereby, from Eq.~(\ref{Eq.2.13}) one gets
\begin{equation}
16J_{1}J_{2}\cos k_{1}\cos k_{2}=0
\end{equation}
which has a solution either $k_{1}$ or $k_{2}=\pi/2$ within the half-BZ. Thus all solutions in this limit lie within the light-cones $x_{1}=2J_{1}t$ and $x_{2}=2J_{2}t$ with speed of light being $2J_{1}a/\hbar$ and $2J_{2}a/\hbar$, where $a$ is lattice spacing. 
\par Now in the gapless phase at $J_{3}=J_{1}=J_{2}$ Eq.~(\ref{Eq.2.12}) again leads to
\begin{equation} \label{Eq.2.14}
\begin{aligned}
x_{1}E^{\prime\prime}_{\mathbf{k}}&=2J^{2}_{1}t\left[\sin k_{1}+\sin (k_{1}+k_{2})\right] \\
x_{2}E^{\prime\prime}_{\mathbf{k}}&=2J^{2}_{1}t\left[\sin k_{2}+\sin (k_{1}+k_{2})\right]
\end{aligned}
\end{equation}
where
\begin{equation}
E_{\mathbf{k}}^{\prime\prime}=J_{1}\left[3+2\cos k_{1}+2\cos k_{2}+2\cos (k_{1}+k_{2})\right]^{1/2}
\end{equation} 
A possible solution to the Eq.~(\ref{Eq.2.14}) could be $k_{1}=k_{2}=\cos^{-1}\sqrt{1-\frac{x_{1}^{2}}{4t^{2}J_{1}^{2}}}=k$ and $x_{1}=x_{2}$. Now Eq.~(\ref{Eq.2.13}) is satisfied for $k=\pi/2$, giving $x_{1}/2J_{1}t=x_{2}/2J_{1}t=\pm 1$. These again corresponds to the light-cones in a sense that $k$ is real if $x_{1}/2J_{1}t$ and $x_{2}/2J_{1}t \leq 1$.
\begin{figure}[h]
\includegraphics[width=\columnwidth]{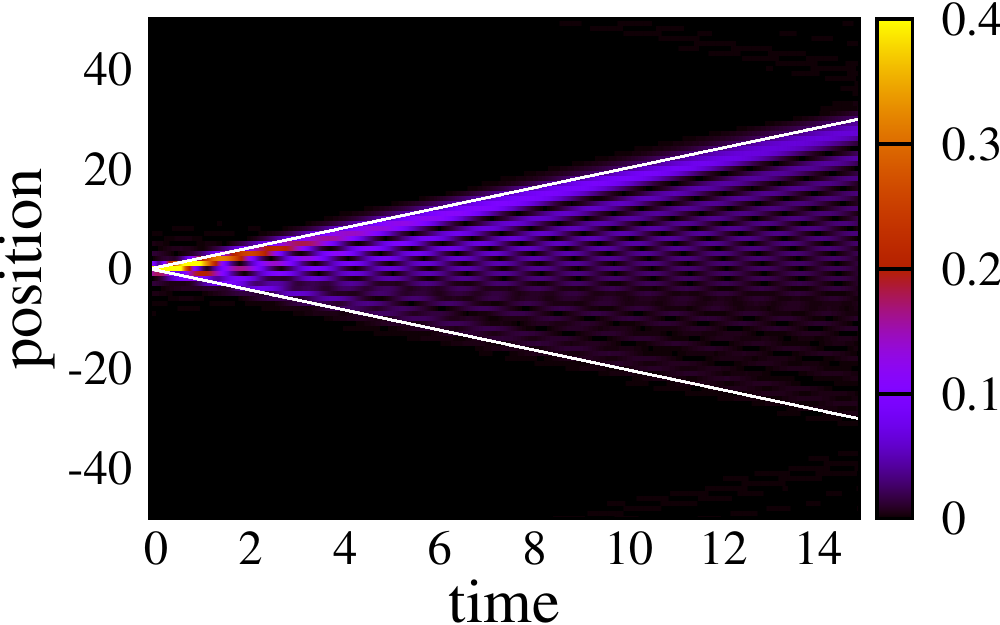}
\caption{(Color online) Plot of $|\psi(\mathbf{r},t)|$ in Eq.~(\ref{Eq.2.9}) depict light-cone like dynamics for $J_{1}=J_{2}=1, J_{3}=3$ and $N=100$. White lines show the envelope representing Lieb-Robinson bound.}
\label{Fig.2.5}
\end{figure}
A plot of the wavefunction amplitude with $x_{1}=x_{2}$ in Fig.~\ref{Fig.2.5} clearly shows the light-cone like dynamics of a Bogoliubov quasiparticle in 2D honeycomb lattice having 100 sites. The two white lines are overplotted using $x_{1}/2J_{1}t=\pm 1$ which perfectly match the boundaries of light-cone.

\section{\label{Sec.3} Floquet evolution}
In this section, we will study what happens when the Hamiltonian varies periodically in time with a period $T$. For numerical purposes \cite{sen16}, let us first consider a protocol where one of the parameters in Kitaev model is given $\delta$-function kicks periodically in time. Here we choose $J_{3}$ in Eq.~(\ref{Eq.2.2}) so that
\begin{equation}
J_{3}(t)=J_{a}+J_{b}\sum_{n=0}^{\infty}\delta(t-nT) 
\end{equation}
where the time period $T$ for one kick is related to the drive frequency as $T=2\pi/\omega$ and $n$ is an integer. We define the Floquet operator $U_{\mathbf{k}}(T,0)$ for each $\mathbf{k}$ at the end of a single period using Eq.~(\ref{Eq.2.5}) as
\begin{equation} \label{Eq.3.1}
\begin{aligned}
U_{\mathbf{k}}(T,0)&=\exp \left(-2iJ_{b}\tau_{3} \right) \exp \left(-2iT(\epsilon_{\mathbf{k}a}\tau_{3}+\Delta_{\mathbf{k}}\tau_{1})\right) \\
\epsilon_{\mathbf{k}a}&=J_{a}+J_{1}\cos k_{1}+J_{2}\cos k_{2} \\
\Delta_{\mathbf{k}}&=J_{1}\sin k_{1}-J_{2}\sin k_{2} 
\end{aligned}
\end{equation}
To proceed further, we note that $U_{\mathbf{k}}$ is a $2\times 2$ matrix which can be written as
\begin{equation}
\begin{aligned}
U_{\mathbf{k}}(T,0)&=\exp \left(-2iJ_{b}\tau_{3}\right) \exp \left(-i(\mathbf{\tau}.\mathbf{n}_{\mathbf{k}})\Phi_{\mathbf{k}a}\right) \\
&=\left( {\begin{array}{cc} A_{\mathbf{k}} & -B_{\mathbf{k}}^{*} \\ B_{\mathbf{k}} & A_{\mathbf{k}}^{*} \\ \end{array}} \right) \\
A_{\mathbf{k}}&=\exp\left(-2iJ_{b}\right)\left(\cos \Phi_{\mathbf{k}a}-in_{3\mathbf{k}}\sin \Phi_{\mathbf{k}a}\right) \\
B_{\mathbf{k}}&=-in_{1\mathbf{k}}\exp\left(-2iJ_{b}\right)\sin \Phi_{\mathbf{k}a}
\end{aligned}
\end{equation}
where $\mathbf{n}_{\mathbf{k}}=\left(\frac{\Delta_{\mathbf{k}}}{E_{\mathbf{k}a}}, 0, \frac{\epsilon_{\mathbf{k}a}}{E_{\mathbf{k}a}}\right)$ and $\Phi_{\mathbf{k}a}=2E_{\mathbf{k}a}T$. Thereby, after $n$ drive cycles, the wavefunction is given by
\begin{equation}
\begin{aligned}
|\psi_{\mathbf{k}}(t=nT)\rangle&=U_{\mathbf{k}}^{n}|\psi_{\mathbf{k}}(t=0)\rangle \\
&=\exp{[-inH_{\mathbf{k}F}T]}|\psi_{\mathbf{k}}(t=0)\rangle
\end{aligned}
\end{equation}
where $H_{\mathbf{k}F}$ is the Floquet Hamiltonian of the system for each wave vector $\mathbf{k}$ \cite{dalessio14,bukov15}, so that $H_{F}=\sum_{\mathbf{k}}H_{\mathbf{k}F}$. To construct $H_{\mathbf{k}F}$, we again use the unitary nature of $U_{\mathbf{k}}$ to express $H_{\mathbf{k}F}$ in terms of the Pauli matrices. This allows us to write
\begin{equation} 
\begin{aligned}
H_{\mathbf{k}F}&=\mathbf{\sigma}.\mathbf{\Lambda}_{\mathbf{k}}=|\mathbf{\epsilon}_{\mathbf{k}F}|(\mathbf{\sigma}.\hat{\Lambda}_{\mathbf{k}}) \\
U_{\mathbf{k}}(T,0)&=\exp{[-iH_{F}T]}=\exp{[-i|\epsilon_{\mathbf{k}F}|(\mathbf{\sigma}.\hat{\Lambda}_{\mathbf{k}})T]}
\end{aligned}
\end{equation}
where $\mathbf{\Lambda}_{\mathbf{k}}=(\Lambda_{1\mathbf{k}}, \Lambda_{2\mathbf{k}}, \Lambda_{3\mathbf{k}})$ and $\hat{\Lambda}_{i\mathbf{k}}=\Lambda_{i\mathbf{k}}/|\mathbf{\epsilon}_{\mathbf{k}F}|$. The quasienergies $\Lambda_{i\mathbf{k}}$ are given by
\begin{equation} \label{Eq.3.2}
\begin{aligned}
\Lambda_{1\mathbf{k}}&=n_{1\mathbf{k}} \sin \Phi_{\mathbf{k}} \cos 2J_{b} \\
\Lambda_{2\mathbf{k}}&=n_{1\mathbf{k}} \sin \Phi_{\mathbf{k}} \sin 2J_{b} \\
\Lambda_{3\mathbf{k}}&=\cos \Phi_{\mathbf{k}} \sin 2J_{b}+n_{3\mathbf{k}} \sin \Phi_{\mathbf{k}}\cos 2J_{b} \\
\end{aligned}
\end{equation}
Here $|\epsilon_{\mathbf{k}F}|=\arccos (M_{\mathbf{k}})/T$ is the Floquet spectrum with
\begin{equation}
M_{\mathbf{k}}=\cos (2J_{b}+\Phi_{\mathbf{k}})+(1-n_{3\mathbf{k}})\sin 2J_{b}\sin \Phi_{\mathbf{k}}
\end{equation}
We therefore write down the wavefunction in position basis as $|\mathbf{r}\rangle=\frac{1}{N}\sum_{\mathbf{k}}\exp{[-i\mathbf{k}.\mathbf{r}]}|\mathbf{k}\rangle$ evolves after one period into
\begin{equation} \label{Eq.3.3}
\psi(\mathbf{r},t)=\frac{1}{N}\sum_{\mathbf{k}}\exp{[-i(\mathbf{k}.\mathbf{r}+|\epsilon_{\mathbf{k}F}|(\mathbf{\sigma}.\hat{\Lambda}_{\mathbf{k}})T)]}
\end{equation}
\begin{figure}[h]
\includegraphics[width=\columnwidth]{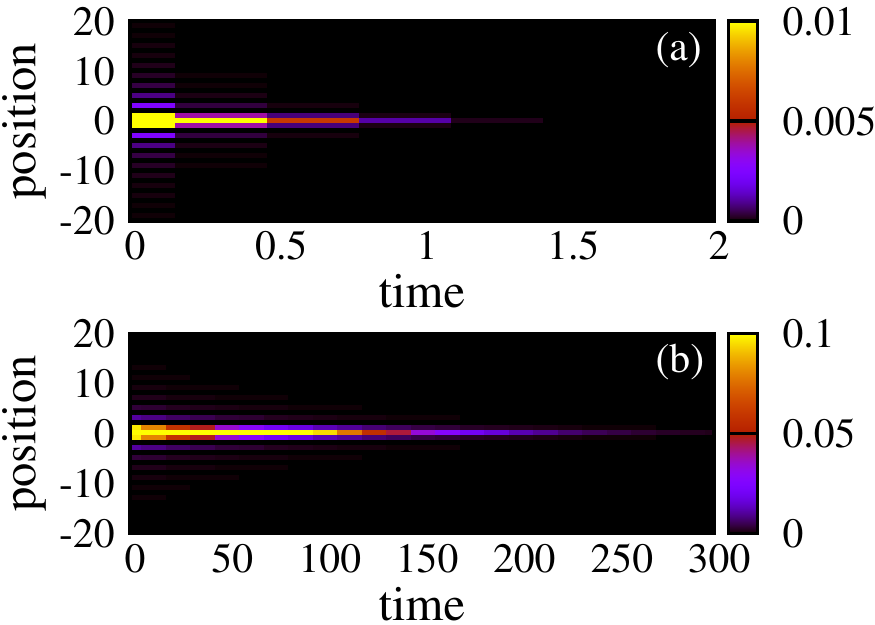}
\caption{(Color online) Plot of $|\psi(\mathbf{r},t)|$ after $\delta$-function kicks over 30 period with (a) $T=0.31$ (b) $T=12.57$. The system being considered has 100 sites and $J_{1}=J_{2}=1, J_{a}=1$ and $J_{b}=0.5$.}
\label{Fig.3.1}
\end{figure}
Using Eq.~(\ref{Eq.3.3}) we compute $|\psi(\mathbf{r},t)|$ with $\delta$-kicked protocol and plot in Fig.~\ref{Fig.3.1} for two different frequency regimes in gapless phase of the model. One can clearly see that the light-cone like spreading of quasiparticle dynamics does not happen in this non-equilibrium case, instead we see some local spreading of wavefunction amplitude at initial times, which decays with each period of driving. This is the effects of stroboscopic observation \cite{kebler12} for typical realizations of this process. The quasiparticle stays confined along $x_{1}=0$ line after each period of observation. The amplitude stays strong for much larger time as $\omega \rightarrow 0$, which suggests that one has to be in this limit in order to observe the long time dynamics. Suitable control on frequency hence necessary for prolonged measurement of quantum information.
\par Next, we consider another drive protocol where $J_{3}(t)$ is varied by squarewave pulse such a way that for a single period of driving $J_{3}$ changes as
\begin{equation}
\begin{aligned}
J_{3}(t)&=J_{a} \quad \text{for} \quad 0 \le t \le T/2 \\
&=J_{b} \quad \text{for} \quad T/2 \le t \le T 
\end{aligned}
\end{equation}
Here one can define the unitary evolution operator for this drive as
\begin{equation}
U_{\mathbf{k}}(T,0)=\exp \left(-iH_{\mathbf{k}a}\frac{T}{2}\right) \exp\left(-iH_{\mathbf{k}b}\frac{T}{2}\right)
\end{equation} 
where $H_{\mathbf{k}a(b)}=\epsilon_{\mathbf{k}a(b)}\tau_{3}+\Delta_{\mathbf{k}}\tau_{1}$ with $\epsilon_{\mathbf{k}a(b)}$ and $\Delta_{\mathbf{k}}$ defined similarly as in Eq.~(\ref{Eq.3.1}). Therefore we get the explicit expressions given by 
\begin{equation} \label{Eq.3.4}
\begin{aligned}
\Lambda_{1\mathbf{k}}&=n_{1\mathbf{k}a} \sin \Phi_{\mathbf{k}a} \cos \Phi_{\mathbf{k}b}+n_{1\mathbf{k}b} \sin \Phi_{\mathbf{k}b} \cos \Phi_{\mathbf{k}a} \\
\Lambda_{2\mathbf{k}}&=\sin \Phi_{\mathbf{k}a} \sin \Phi_{\mathbf{k}b} \left(n_{1\mathbf{k}a}n_{3\mathbf{k}b}-n_{1\mathbf{k}b}n_{3\mathbf{k}a}\right) \\
\Lambda_{3\mathbf{k}}&=n_{3\mathbf{k}a} \sin \Phi_{\mathbf{k}a} \cos \Phi_{\mathbf{k}b}+n_{3\mathbf{k}b} \sin \Phi_{\mathbf{k}b} \cos \Phi_{\mathbf{k}a} \\
M_{\mathbf{k}}&=\cos \Phi_{\mathbf{k}a} \cos \Phi_{\mathbf{k}b}-n_{\mathbf{k}a}.n_{\mathbf{k}b} \sin \Phi_{\mathbf{k}a} \sin \Phi_{\mathbf{k}b}
\end{aligned}
\end{equation}
\begin{figure}[h]
\includegraphics[width=\columnwidth]{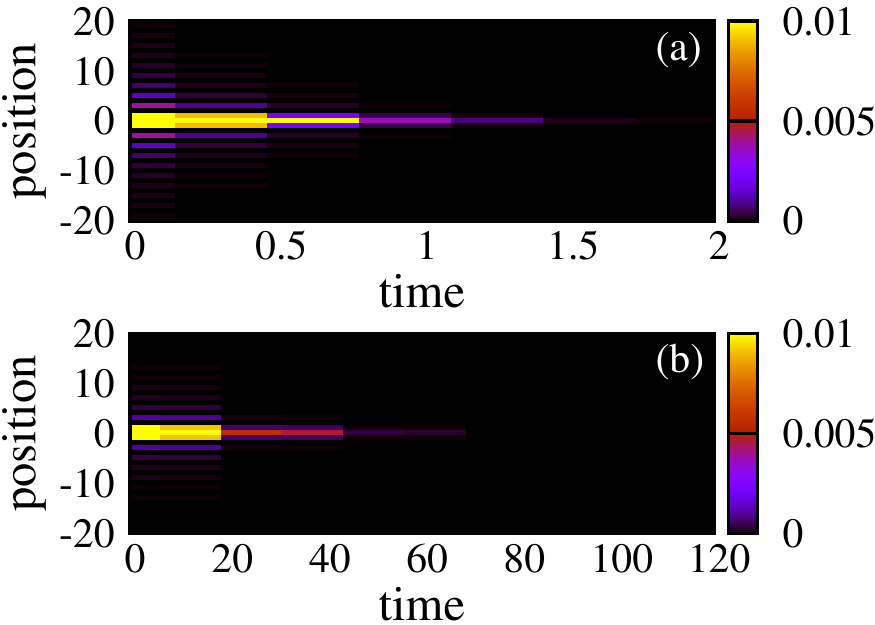}
\caption{(Color online) Plot of $|\psi(\mathbf{r},t)|$ for square pulse protocol. All parameters have kept same as in Fig.~\ref{Fig.3.1}.}
\label{Fig.3.2}
\end{figure}
We plot again $|\psi(\mathbf{r},t)|$ in Fig.~\ref{Fig.3.2} considering square pulse protocol and show similar results except the fact that, in this case, the dynamics persists less time compared to $\delta$-kicked protocol for same drive parameters ($n$ and $T$). Here we note that the above results for two independent drive protocols also hold for gapped phase.
\section{\label{Sec.4} Discussion}
In this paper, we have showed the existance of light-cone like dynamics of quasiparticle propagation as a signeture of quantum caustics in 2D Kitaev honeycomb model. We have found that this dynamics happens at a specific direction in lattice plane and exhibits spatial anisotropy depending on model parameters. This directional behaviour is a remarkable feature in a sense that all the information travels at a specific angle at equilibrium. This has not been an issue for previous 1D systems, thus plays a key role in experimental measurement of quantum information in 2D. We have applied conditions for stationary action to get exact analytic expressions to the caustics envelops giving the maximum speed limit for quantum information propagation in this system. We have introcuced external time dependent drive protocols by periodically varying one of the model parameters with a specific time period controlled by drive frequency. For two independent drive protocols, we have showed that the caustics no longer follow the light-cone like structure and lead to the localization of quasiparticle.
\par There have been proposals for realizing Kitaev model using ultracold atoms and molecules trapped in optical lattice systems \cite{duan03,micheli06}. If this can be done, one can trace the propagation of quasiparticle through the 2D honeycomb lattice. Using external laser pulses or electric field, one can even experimentally localize a single quasiparticle and get control over its motion by simply tuning appropriate drive frequency.

\section*{Acknowledgements}
Author thanks K. Sengupta for discussions and acknowledges Department of Science and Technology, Government of India for financial support through INSPIRE program.

\bibliographystyle{apsrev4-1}
\bibliography{kitaevcaustics.bib}

\end{document}